\documentclass[useAMS,usenatbib]{mn2e}
\usepackage{graphicx}
\usepackage{amssymb}

\newcommand{\mc}{\multicolumn}
\newcommand{\beqn}{\begin{equation}}
\newcommand{\eeqn}{\end{equation}}

\def\tablenotemark#1{{$^#1$}}

\def\kms{\,{\rm km\,s^{-1}}}
\def\Gyr{\,{\rm Gyr}}

\def\kpc{\,{\rm kpc}}

\def\rp{r_{\rm p}}
\def\dm{\delta m_{\rm r}}
\def\dvr{\delta v_{\rm r}}
\def\tmerge{t_{\rm merge}}
\def\fmass{f_{\rm mass}}

\begin{document}

\title[Major Dry-Mergers In BCGs]
{Major Dry-Mergers In Early-Type Brightest Cluster Galaxies}
\author[F. S. Liu et al.]{
F. S. Liu$^{1,2}$\thanks{E-mail: lfs@bao.ac.cn},
Shude Mao$^{3}$,
Z. G. Deng$^{4}$,
X. Y. Xia$^{5}$,
and Z. L. Wen$^{2}$
\\
$^{1}$College of Physics Science and Technology,
Shenyang Normal University, Shenyang, 110034, P.R.China\\
$^{2}$National Astronomical Observatories, Chinese
Academy of Sciences, A20 Datun Road, Beijing, 100012, P.R.China\\
$^{3}$Jodrell Bank Centre for Astrophysics, Alan Turing Building, University of Manchester,
Manchester, M13 9PL, UK\\
$^{4}$Graduate University of Chinese Academy of Sciences,
                 Beijing, 100049, P.R.China\\
$^{5}$Tianjin Astrophysics Center, Tianjin Normal University, Tianjin, 300074, P.R.China\\
}

\date{Accepted 2009 April 14, Received 2009 April 5; in original form 2009 January 27}

\pagerange{\pageref{firstpage}--\pageref{lastpage}} \pubyear{2009}

\maketitle

\label{firstpage}

\begin{abstract}

We search for ongoing major dry-mergers in a well selected sample
of local Brightest Cluster Galaxies (BCGs) from the C4 cluster
catalogue. 18 out of 515 early-type BCGs with redshift between
0.03 and 0.12 are found to be in major dry-mergers, 
which are selected as pairs (or triples) with $r$-band magnitude difference
$\dm<1.5$ and projected separation $\rp<30$ kpc, and showing 
signatures of interaction in the form of significant asymmetry in residual images. 
We find that the fraction of BCGs in major dry-mergers increases with the richness
of the clusters, consistent with the fact that richer clusters
usually have more massive (or luminous) BCGs. We estimate that
present-day early-type BCGs may have experienced on average $\sim
0.6\,(\tmerge/0.3\Gyr)^{-1}$ major dry-mergers and through this
process increases their luminosity (mass) by $15\%\,
(\tmerge/0.3\Gyr)^{-1} \,(\fmass/0.5)$ on average since $z=0.7$,
where $\tmerge$ is the merging timescale and $\fmass$ is the mean
mass fraction of companion galaxies added to the central ones. We
also find that major dry-mergers do not seem to elevate radio
activities in BCGs. Our study shows that major dry-mergers
involving BCGs in clusters of galaxies are not rare in the local
Universe, and they are an important channel for the formation and
evolution of BCGs.
\end{abstract}


\section{Introduction}

The Brightest Cluster Galaxies (BCGs) are at the most luminous and
massive end of galaxy population. They are usually located at or
close to the centres of dense clusters of galaxies based on the X-ray
observations or gravitational lensing observations (e.g., Jones \&
Forman 1984; Smith et al. 2005). Most of them
are dominated by old stars without prominent ongoing star
formation. They are much more luminous than other galaxies in the
clusters, and have distinct surface brightness profiles from other
galaxies (see Vale \& Ostriker 2008 and references therein). And
yet the luminosities of BCGs appear to have rather small scatters.
As a result, they have been proposed as a distance indicator (e.g.
Postman \& Lauer 1995). Because of the unusual properties of BCGs,
their formation and evolution has long attracted particular
attention.

Galactic cannibalism as a way to build up the BCGs was proposed
more than three decades ago (e.g., Ostriker \& Tremaine 1975;
White 1976; Vale \& Ostriker 2008). However, as pointed out by
several authors (e.g., Merritt 1985; Lauer 1988; Tremaine 1990),
cannibalism of satellite galaxies may not be enough to account for
the luminosity growth of massive BCGs due to the small luminosity
ingestion rate. If this is correct, then other mechanisms, such as
major mergers may play an important role (Lin \& Mohr 2004; Vale
\& Ostriker 2008). Since most central galaxies in clusters have
little cold gas, the mergers are not expected to trigger major
episodes of star formation and thus they are likely dry mergers.

Recent studies from both numerical simulations and observations
indicate that a large part of stellar mass in luminous early-type
galaxies form before redshift $z \sim 1-2$, and later dry mergers
play an important role in their stellar mass assembly (Gao et al.
2004; De Lucia \& Blaizot 2007; van Dokkum et al. 1999; Tran et
al. 2005). Ruszkowski \& Springel (2009) investigated the effect of dry-mergers
on the scaling relations of BCGs in simulations.
Observational studies on the properties of luminous
early-type galaxies provide evidence that they have experienced
dry-mergers (e.g., Lauer et al. 2007; Bernardi et al. 2007; von
der Linden et al. 2007; Liu et al. 2008, hereafter Paper I).
Theoretical investigations suggest that half the mass of a typical
BCG may be assembled via accretion of smaller galaxies, i.e., by
minor mergers at $z<$ 0.5 (e.g., Gao et al. 2004; De Lucia \&
Blaizot 2007). Recently, Bernardi (2009) studied 
their evolutions in the size- and velocity dispersion-stellar mass
correlations since $z<0.3$, and concluded that early-type BCGs grow from
many dry minor mergers (see Discussion). However, Whiley et al. (2008) 
using the ESO Distant Cluster Survey, found little evolution of BCGs
since $z\sim1$, at least within a metric circular aperture of 37\,kpc.
A consensus about the evolution of BCGs is yet to emerge. 

Recent modelling suggests that major dry-mergers play a
significant role in forming luminous, intermediate-mass,
early-type galaxies (e.g. Naab et al. 2006) at redshift $z \leq
0.5$. Whether the same scenario applies to the much more luminous
BCGs, and more generally how the occurrence of major dry-mergers
depends on the environments, is still somewhat controversial.
Merritt (1985) argued that massive major mergers are more likely
to occur in large groups than in massive clusters, since the
merger rate is a steeply declining function of the velocity
dispersions of a virilized system.

However, recent observational studies indicate major dry-mergers
do occur in groups and clusters of galaxies. For example, 
Mulchaey et al. (2006) and Jeltema et al. (2007) reported some examples of
dry-mergers involving central galaxies in intermediate-redshift groups.
Tran et al. (2008) reported an observational analysis of supergroup SG
1120-1202 at $z \sim 0.37$, which is expected to merge and form a
cluster with mass comparable to Coma. They argued that the group
environment is critical for the process of major dry-mergers.
Rines et al. (2007) reported a very massive cluster (CL0958+4702)
at moderate redshift ($z=0.39$), in which a major dry-merger is
ongoing to build up a BCG. Using the Sloan Digital Sky Survey (SDSS) data, 
McIntosh et al. (2008) identified 38 major mergers of red galaxies from 845 
groups of galaxies at $z \leq 0.12$ identified
using the halo-based group finder of Yang et al. (2005),
and showed that centres of massive groups are the preferred environment for major
dry-mergers and mass assembly of massive early-type galaxies.

All these studies show that major dry-mergers are important in the
formation of BCGs at moderate redshifts. Our own visual
inspections of 249 merging pairs identified from luminous
early-type galaxies (Wen, Liu \& Han 2009) also hint that some of
these mergers involve BCGs. The purpose of this work is to perform
a search for major dry-mergers in a well-selected local early-type
BCG sample with a quantitative method and then estimate their
major dry-merger and luminosity (mass) increase rates. We also investigate the
dependence of the merger fraction as a function of the richness of
the clusters.

The structure of the paper is as follows. In \S2 and \S3 we
describe our sample and identifications of major dry-mergers
involving BCGs. We present our main results in \S4 and finish
with a summary and discussion in \S5. Throughout this paper we
adopt a cosmology with a matter density parameter $\Omega_{\rm
m}=0.3$, a cosmological constant $\Omega_{\rm\Lambda}= 0.7$, and a
Hubble constant of $H_0=72\,{\rm km \,
  s^{-1} Mpc^{-1}}$, i.e., $h=H_0/(100\,{\rm km \,  s^{-1} Mpc^{-1}})=0.72$.

\section{The Sample\label{sec:sample}}

The C4 cluster catalogue (Miller et al. 2005) was constructed from the 
SDSS spectroscopic data in the
parameter space of position, redshift, and color. The C4 algorithm identifies 
the BCG from the spectroscopic catalogue within $0.5 h^{-1}$\,Mpc of 
the central galaxy at the peak of the C4 density field (``mean'' galaxy). 
The early (DR2) version of the C4 catalogue incorporates the SDSS 
photometric catalogue to select the brightest cluster galaxy within $1 h^{-1}$\,Mpc 
in order to avoid missing BCGs in spectroscopic data due to fiber collisions.  
The catalogue is relatively complete in the redshift range $0.03 \leq z \leq 0.12$, 
including 98\% of X-ray-identified clusters and 90$\%$ of Abell clusters in the same
region (Miller et al. 2005). We thus limit our data within this
range of redshift, obtaining 643 clusters after we reject 16
duplicates (Bernardi et al. 2007).

\begin{figure}
\centering
\includegraphics[width = 8.4cm]{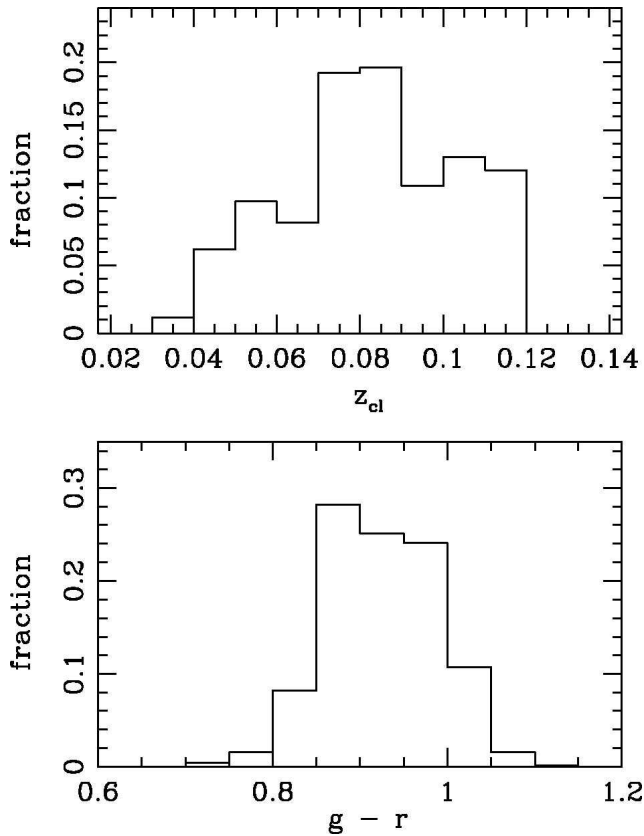}
\caption{The top panel shows the redshift distribution of redshift $z_{cl}$ 
for the 515 C4 clusters while the bottom panel shows the colour $g - r$ distribution of
their early-type BCGs.} \label{fig:samp}
\end{figure}

However, as pointed out by several previous works (e.g., Bernardi
et al. 2007; von der Linden et al. 2007), some bright stars
have been mis-classified as BCGs, and about 1/4 of BCGs show
late-type features (e.g., distinct spiral arms, dust features). 
We thus re-checked each cluster and identified 
the brightest galaxy within $1 h^{-1}$\,Mpc of the cluster center as the BCG. 
We then select a sample of early-type BCGs from this catalogue by 
excluding artifact stars and late-type galaxies.  
Furthermore we require the BCGs should be
in clusters with richness \footnote{The richness is defined as the
number of member galaxies within $1 h^{-1}$\,Mpc of the central galaxy.} larger than 10, and
have colour typical of early-type galaxies, $g-r>0.7$ (Strateva et
al. 2001) in the SDSS model magnitudes. In the end we obtain a
sample of 515 early-type BCGs in the local Universe.
Figure~\ref{fig:samp} shows the distributions of redshifts of the
clusters and $g-r$ colours of BCGs, which has taken into account
the extinction by the SDSS, and k-correction to $z=0.1$ (Blanton
\& Roweis 2007). The cluster redshifts range from 0.03 to 0.12,
with a maximum around 0.08; the $g-r$ colours of BCGs peak around
0.9 with a small RMS scatter of around 0.06, consistent with their
stellar populations being quite old and uniform.

\section{Identifications of major dry-mergers involving BCGs}\label{iden}

Dry-mergers between gas-poor early-type galaxies are generally
more difficult to identify than those mergers involving late-type
galaxies (Naab et al. 2006). The latter often develop prominent
tidal tails dotted with star-forming regions. However, dry-mergers
have more subtle features, such as broad ``fans'' due to ejected
stars, asymmetries in inner isophotes and/or sometimes diffuse
tails (Rix $\&$ White 1989; Combes et al. 1995). While major
dry-mergers usually have more prominent morphological signatures
than minor dry-mergers (Le F\`evre et al. 2000), they still
exhibit much weaker merging features than wet-mergers.

\begin{figure}
\hspace{-0.8cm}
\includegraphics[width=1.2\columnwidth]{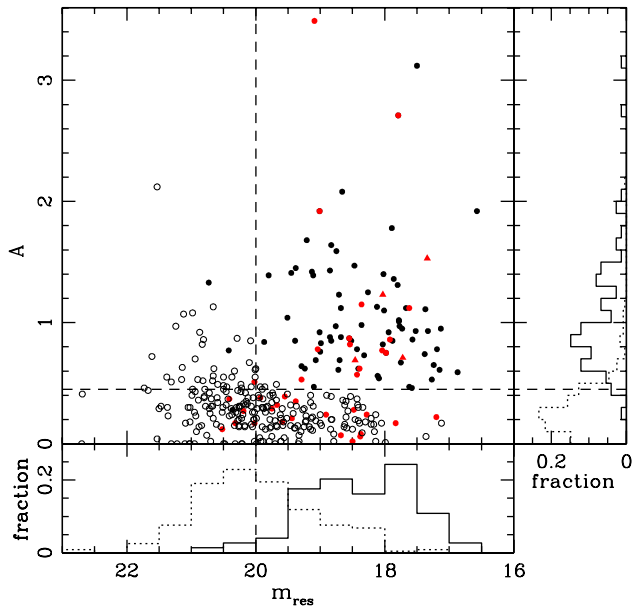}
\caption{The $r-$band residual magnitudes $m_{\rm res}$ versus the asymmetry
factor ${\cal A}$. The black points are the pairs in Wen et al. (2009,
Fig. 5 in their paper). The 36 merger candidates with $7 <\rp<30 \kpc$
are shown with red circles. Four additional mergers with $\rp<7 \kpc$ are shown
with red triangles. The horizontal and vertical dotted lines
illustrate the criteria of Wen et al. e.g., $m_{\rm res}<20$\,mag
and ${\cal A}>0.45$ used to select major mergers. The histograms
for ${\cal A}$ and $m_{\rm res}$ for the Wen et al. (2009) sample are shown at the right and
bottom for major-mergers with strong interactions (solid histogram) and those classified as weak
interactions or due to chance alignment (dashed histogram).
} \label{fig:diagnose}
\end{figure}

We adopt the following criteria to select candidates of ongoing
major dry-mergers involving the BCG in our sample. First, galaxies
with projected distance from central BCG must satisfy $\rp<30
\kpc$. We search for close pairs both automatically and visually.
In the automated search, a lower limit of the projected distance,
7 \kpc, is imposed, which corresponds to a separation of $3''$ at
$z=0.12$. This restriction is necessary because the photometry
becomes unreliable for galaxies with smaller angular separation
(McIntosh et al. 2008; Wen, Liu \& Han 2009). Pairs or triples
with separations smaller than $7\kpc$ are searched visually.
Second, they should have magnitude difference with their
corresponding central BCGs $<2$ magnitudes in the SDSS $r-$band
model magnitude.  The magnitude difference applied here is
slightly larger than the value (1.5) that will be eventually used to select
major-mergers. The reason is that the SDSS magnitudes for BCGs are
somewhat inaccurate due to the problem of sky background
subtraction (Paper I). Third, both galaxies in a pair must have
$g-r>0.7$ to ensure the merging galaxies are early-type (Strateva
et al. 2001). In total, there are 49 apparent close pair
candidates at $7 <\rp< 30 \kpc$ satisfying these three conditions (pairs
with $\rp<7$\,kpc are discussed below).

Most previous studies impose an additional requirement of a small
difference in the line-of-sight velocities $\dvr$ (e.g., $\dvr <
500 \kms$) besides a small projected distance $\rp$ to avoid
chance alignment. However, SDSS spectroscopic survey is severely
incomplete on very small angular scales due to fiber collision
(McIntosh et al. 2008; Wen, Liu \& Han 2009). Hence it is not
possible to apply this criterion to the SDSS data. An alternative 
method is needed to identify merging systems from the projected pairs and triples.

Although, as mentioned above, the interaction features (e.g.,
broad plumes at the outskirts, short tidal tails, bridges or
asymmetries  in inner isophotes) are weak in mergers of early-type
galaxies, they can nevertheless be seen more clearly from the
residual image after we subtract a smooth and symmetric model for
each galaxy in an image (Bell et al. 2006; McIntosh et al. 2008;
Wen, Liu \& Han 2009). Such a procedure has been described in
detail by Wen, Liu \& Han (2009). For completeness, we repeat the
essentials of the method below.

First, we extract the SDSS $r-$band images of all 49 close pairs
and triples. The sky backgrounds are subtracted precisely from the
corrected frames with the method of Paper I. Briefly, the background 
is subtracted as follows: all the objects including target sources 
in the corrected frame were firstly masked to obtain a background-only image. 
Second-order Legendre polynomials are then used to fit the smoothed background-only 
image to obtain an accurate sky background model. After background subtraction, 
the GALFIT package (Peng et al. 2002) is then used to construct a smooth symmetric
S{\'e}rsic (S{\'e}rsic 1968) model for every early-type galaxy in
the projected pair or triple. Stars and fainter galaxies with
distances closer than $2R_{90}$ from the centre of BCGs (Here
$R_{90}$ is the radius containing 90\% of Petrosian flux from the
SDSS catalogue) are modelled with a double-Gaussian
point-spread-function (Stoughton et al. 2002) and S{\'e}rsic
function respectively. Objects outside $2R_{90}$ in the extracted
images are masked. All models are convolved with the
point-spread-function; we then obtain the best model by minimizing
the $\chi^2$ with the sky-subtracted image of unmasked pixels.

A fitted magnitude for each target galaxy and a residual image for
each pair are obtained afterwards. Notice that the fitted
magnitude is superior because it can separate the flux in the
overlapping region of pairs and correct the over-estimate in the
sky background in the SDSS pipeline (Paper I). 
Wen, Liu \& Han (2009) improved the method of Conselice et al. (2000) 
to calculate the asymmetry factor for residual images of galaxy pairs. 
Essentially it measures the difference between pixels and those symmetric 
pixels with respect to the centre and major axis of each galaxy (see eqs. 8-10 in 
their paper). The pixels in overlapping regions are treated separately
(see their Figure 4). The asymmetry factor ${\cal A} \sim 0$ is for a galaxy pair 
without any interaction feature ideally. A large ${\cal A}$ means a stronger asymmetric interaction.
In order to find an efficient criterion, they compared the asymmetry factors 
of two test samples of residual images selected through careful visual inspections: a sample with 
distinct interaction features and another almost without. Both are not contaminated 
by other objects within $2R_{90}$ of target galaxies. The test showed that a close pair can 
be classified efficiently as a merging system with distinct interaction features when the residual 
images have an $r-$band magnitude $r_{\rm res}<20$~mag and an asymmetry factor ${\cal A} > 0.45$. 
In this work, we performed further extensive tests and find that both
the asymmetry parameter ${\cal A} > 0.5$ and residual magnitude $r_{\rm res}<19.5$ can select strongly
interacting mergers efficiently (see the two histograms in Figure~\ref{fig:diagnose}). These
criteria are only slightly different from those used by Wen, Liu \& Han (2009). 
The criteria of $r_{\rm res}<19.5$~mag is almost the same as 
requiring the flux ratio of the residual image to the initial image to be larger than $2\% $
($f_{\rm res}/f_{\rm ini}>2\%$, see the last column in Table~\ref{tab:bcg}).

\begin{figure*}
\centering
\includegraphics[width = 17.8cm]{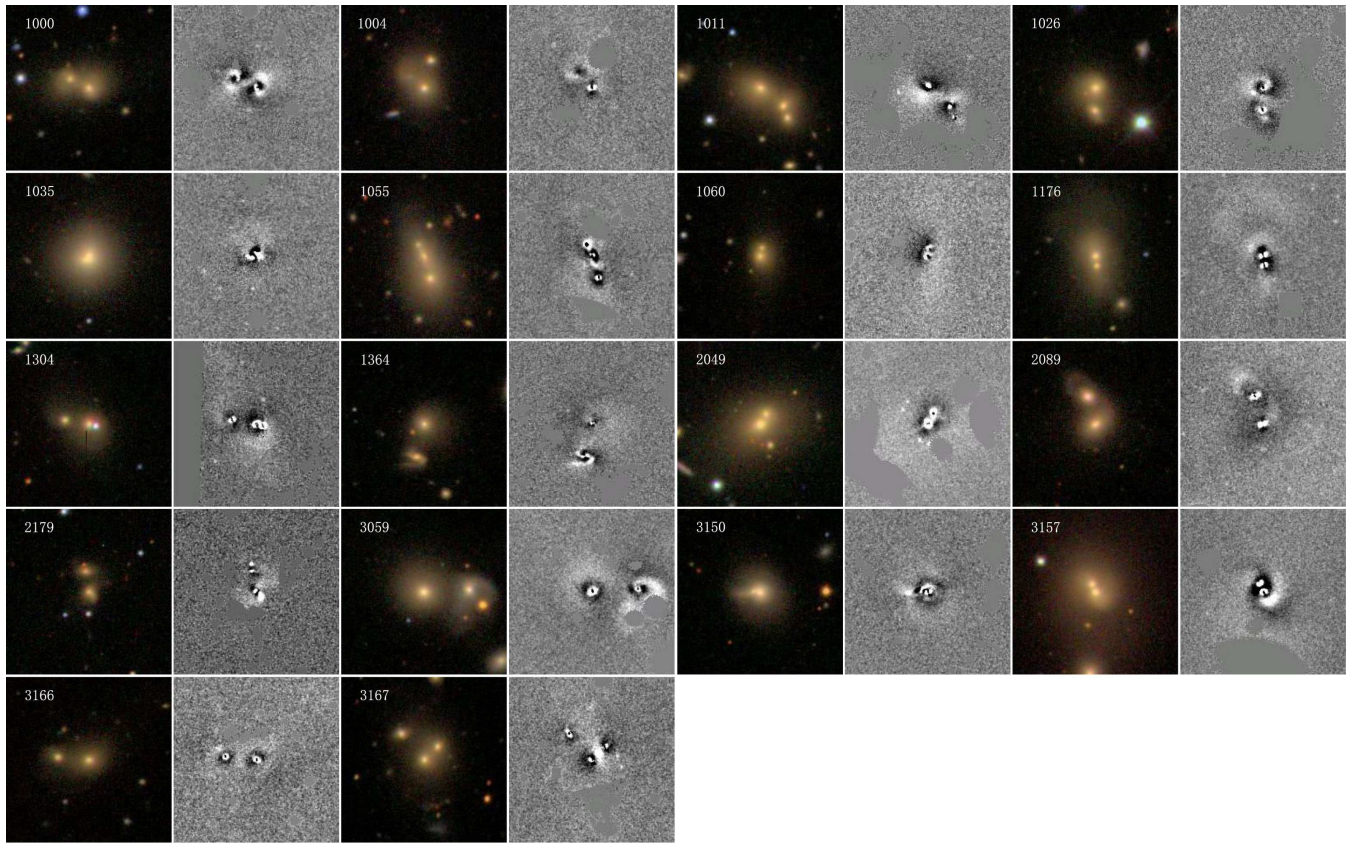}
\caption{
The colour images and corresponding residual images in
the $r-$band for the 18 identified  major mergers. Each image is
$80'' \times 80''$. The SDSS-C4 name for the DR2 version is marked
at the top left corner of each colour image.
} \label{fig:merger}
\end{figure*}

\begin{figure}
\centering
\includegraphics[width = 8.5 cm]{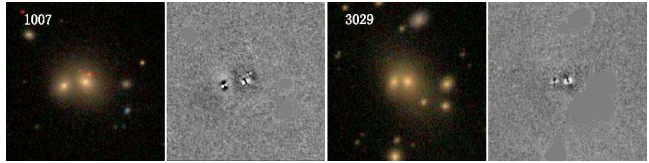}
\caption{Two typical example galaxies which after subtraction show small residuals and small asymmetry
  factors. They are likely pairs produced by either weakly interacting with each
  other or chance alignment. Each image is $80'' \times 80''$. } \label{fig:nomerg}
\end{figure}

If we limit major mergers as having a fitted magnitude difference
$\dm<$ 1.5, we find that 36 out of 49 close pairs or triples
satisfy 7 $<\rp<$ 30 \kpc. Furthermore, 14 out of 36 have
residuals $r_{\rm res}< 19.5$~mag and asymmetry factors ${\cal A}
> 0.5$ (see Figure~\ref{fig:diagnose}). Others may be weak interacting 
systems or pairs in chance alignment; two examples with small residuals
are shown in Figure~\ref{fig:nomerg}.
We added 4 mergers (C1035, C1176, C3150, C3157) with $\rp<$ 7
\kpc~(e.g., mergers with close double nuclei) identified visually
-- these systems have fiber magnitude difference $\dm<1.0$, and
can be classified as merging systems if we apply the same criteria
as mergers at $7 <\rp< 30 \kpc$. The $r-$band observational images
and corresponding residual images for these 18 mergers involving
BCGs are shown in Figure~\ref{fig:merger}. Their corresponding
parameters are listed in Table~\ref{tab:bcg}.

Bernardi et al. (2003) derived the luminosity function of
early-type galaxies. In the $r$-band, the magnitude for an $L_*$
galaxy (derived from a Gaussian fit) 
is $M_*=-21.09$ (taking $h=0.72$ instead of their $h=0.70$) . The BCGs in our sample have luminosities
ranging from $L_*$ to roughly $\sim 10L_*$, with a mean luminosity
of $\sim5L_*$. For C4 2089, the merging pair has luminosities
$11.2L_*$ and $5.6 L_*$. Similarly, for C4 1055, the triple merging
galaxies have luminosities $7.9 L_*, 6.6 L_*$ and $0.46L_*$; the
two most luminous components have identical radial velocities, and
thus there is little doubt that they are physically associated.
Mergers appear to occur even among the most luminous galaxies in
clusters (see \S\ref{sec:rich}). If all the stars coalesce to form
a single giant galaxy, then the luminosities will be as high as
$16.8 L_*$ and $15L_*$; such galaxies are rarely seen. This
suggests that a substantial amount of stars of the companion are
tidally stripped to form an ``envelope'' around the central
galaxy, a point we return to in \S\ref{sec:assembly}.

\begin{table*}
\caption[]{Basic parameters for the 18 identified major dry-mergers involving BCGs}
\begin{center}
\begin{tabular}{lccrrccclccrrc}
\hline
\mc{1}{r}{C4 ID} & \mc{1}{c}{$z_{\rm mg}$} & \mc{1}{r}{$M_{tot}$} &
\mc{1}{r}{R.A.(J2000)} & \mc{1}{c}{Dec.(J2000)} & \mc{1}{c}{$z_{\rm
    sp}$} & \mc{1}{c}{$\rp$} & \mc{1}{c}{$\dvr$} & \mc{1}{c}{$m_r$}
& \mc{1}{c}{$M_r$}  & \mc{1}{c}{$\tmerge$} & \mc{1}{c}{$A$} & \mc{1}{c}{$m_{\rm res}$} &
\mc{1}{c}{${f_{\rm res}/f_{\rm ini}}$}\\

\mc{1}{r}{} & \mc{1}{r}{} & \mc{1}{r}{(mag)} & \mc{1}{c}{} & \mc{1}{c}{} & \mc{1}{l}{} & \mc{1}{c}{(\kpc)} & \mc{1}{c}{($\kms$)} & \mc{1}{c}{(mag)} & \mc{1}{c}{(mag)} & \mc{1}{c}{($\Gyr$)} & \mc{1}{c}{} & \mc{1}{c}{(mag)} & \mc{1}{c}{(\%)} \\

\mc{1}{c}{(1)} & \mc{1}{c}{(2)} & \mc{1}{c}{(3)} & \mc{1}{c}{(4)} & \mc{1}{c}{(5)} & \mc{1}{c}{(6)} & \mc{1}{c}{(7)} & \mc{1}{c}{(8)} & \mc{1}{c}{(9)} & \mc{1}{c}{(10)} & \mc{1}{c}{(11)} & \mc{1}{c}{(12)} & \mc{1}{c}{(13)}& \mc{1}{c}{(14)} \\

\hline
1000&0.0864&-23.36&202.543025 &-2.105012&  0.0866&      &       & 15.42&-22.62  & 0.32 & 0.75& 17.98 & 5.2\\
    &      &      &202.545566 &-2.103784&  0.0863& 15.94&  90   & 15.44&-22.60  &     &      & \\
1004&0.0814&-23.30&149.717402 & 1.059187&  0.0814&      &       & 14.90&-22.99  & 0.27 & 0.78& 19.04& 2.5\\
    &      &      &149.719361 & 1.060678&      --& 13.18&--& 16.09&-21.80  &      &      & \\
1011&0.0903&-23.91&227.107326 &-0.266266&  0.0903&      &       & 14.87&-23.27 & 0.39 &  2.71& 17.79 & 5.1\\
    &      &      &227.104246 &-0.268636& --& 22.92&--& 15.51&-22.63  &      &      & \\
    &      &      &227.103726 &-0.270214& --& 31.50&--& 16.38&-21.76  &      &      & \\
1026&0.0908&-24.16&191.926938 &-0.137254& --&      &       & 14.50&-23.65  & 0.28 &  1.15& 18.36 & 2.3\\
    &      &      &191.927005 &-0.140256&  0.0908& 17.78&--& 15.06&-23.09  &      &      & \\
1035$^*$&0.0444&-22.95&175.872249 &-1.667968&  0.0446&      &       & 16.11\tablenotemark{a}&-- & 0.04 &  0.69& 18.46& 2.1\\
    &      &      &175.872512 &-1.668436&  0.0442&  1.64& 120   & 16.16\tablenotemark{a}&-- &     &   & \\
1055&0.0836&-24.03&202.795956 &-1.727285&  0.0836&      &       & 14.61&-23.34  & 0.28 &  0.62& 18.39 & 2.1\\
    &      &      &202.795126 &-1.730259&  0.0836& 16.99&  0    & 14.81&-23.14  &     &      & \\
    &      &      &202.796700 &-1.725844& --&  8.93&--& 17.70&-20.25  &      &      & \\
1060&0.1170&-23.29&212.497855 &-1.539655& --&      &       & 15.87&-22.91  & 0.17 &  3.49& 19.09 & 4.2\\
    &      &      &212.497645 &-1.538576&  0.1170&  8.13&--& 16.81&-21.97  &      &      & \\
1176$^*$&0.0737&-23.90&189.734808 & 6.158386&  0.0738&      &       & 17.06\tablenotemark{a}&-- & 0.11 &  1.53& 17.34& 3.3\\
    &      &      &189.734648 & 6.157096&  0.0736&  6.37&  60   & 17.62\tablenotemark{a}&-- &      &  &  \\
1304&0.0986&-22.69&148.914494 & 1.596777&  0.0987&      &       & 16.12&-22.23  & 0.50 &  0.87& 18.55 & 7.4\\
    &      &      &148.917693 & 1.597327&  0.0986& 20.72&  30   & 16.84&-21.51  &      &    &  \\
1364&0.0708&-22.75&154.700071 & 0.385141&  0.0710&      &       & 15.06&-22.49  & 0.50 &  0.82& 18.54 & 4.9\\
    &      &      &154.701081 & 0.380757&  0.0707& 21.27&  90   & 16.49&-21.06  &      &     & \\
2049&0.0751&-23.89& 16.841087 &14.273220& --&      &       & 14.05&-23.64  & 0.13 &  0.77& 18.04 & 2.2\\
    &      &      & 16.840364 &14.274536&  0.0751&  7.44&--& 15.52&-22.17  &      &     & \\
2089&0.0652&-24.15& 59.582654 &-5.538873&  0.0652&      &       & 13.65&-23.71 & 0.27 &  0.86& 17.92 & 5.1\\
    &      &      & 59.583713 &-5.535151& --& 16.95&--& 14.40&-22.96  &      &    &  \\
2179&0.0938&-22.24&322.027369 &11.405484& --&      &       & 16.71&-21.53  & 0.47 &  0.57& 18.43 & 10.6\\
    &      &      &322.027823 &11.408263&  0.0938& 17.17&--& 16.80&-21.44  &      &    &   \\
3059&0.0406&-22.47&122.148487 &38.914504& --&      &       & 14.26&-22.00  & 0.43 &  1.12& 17.62 & 4.4\\
    &      &      &122.140730 &38.914841&  0.0406& 16.97&--& 14.94&-21.32  &      &    & \\
3150$^*$&0.0456&-22.05&134.299844 &53.470904&  0.0456&      &       & 16.43\tablenotemark{a}&-- & 0.06 &  0.71& 17.72& 4.8\\
    &      &      &134.301020 &53.470989& --&  2.21&--& 17.09\tablenotemark{a}&-- &    &    & \\
3157$^*$&0.0635&-24.10&176.404931 &64.511485&  0.0632&      &       & 16.37\tablenotemark{a}&-- & 0.08 &  1.23& 18.03& 2.3\\
    &      &      &176.406011 &64.512554&  0.0638&  4.97& 180   & 16.89\tablenotemark{a}&-- &      &   &  \\
3166&0.1112&-24.04&192.035328 &64.036916&  0.1112&      &       & 14.93&-23.72  & 0.49 &  0.53& 19.29 & 3.3\\
    &      &      &192.044829 &64.037322& --& 29.63&--& 16.11&-22.54  &      &    &  \\
3167&0.0915&-23.36&256.010966 &33.868847&  0.0915&      &       & 15.88&-22.29  & 0.55 &  1.92& 19.01 & 3.5\\
    &      &      &256.014668 &33.872302&  0.0919& 27.66& 120   & 15.94&-22.23  &      &    & \\
    &      &      &256.008867 &33.870576&  0.0911& 14.55&  90   & 16.25&-21.92  &      &    & \\
\hline
\end{tabular}
\end{center}
{Note:
Col:(1). The C4 cluster ID (DR2) for the Clusters. The four
visually-selected ones (indicated by $*$) have $\rp<7\kpc$ (see \S\ref{iden}).
Col:(2). Redshifts of identified merging pairs (or triples).
Col:(3). Total absolute magnitude of mergers in the SDSS $r-$band.
Col:(4). R.A. (J2000) of the component galaxy in a merger.
Col:(5). Dec. (J2000) of the component galaxy in a merger.
Col:(6). Spectroscopic redshift of the component galaxy in a merger.
Col:(7). The projected distance, $\rp$.
Col:(8). The line-of-sight velocity difference, $\dvr$, if available.
Col:(9). Extinction-corrected $r-$band fitted apparent magnitude of
component galaxy in a merger. For four mergers with $\rp<$ 7 \kpc~
(marked with `a'), we compare their fiber magnitudes.
Col:(10). $r-$band absolute magnitude of component galaxy in a
merger, corrected for extinction (by SDSS) and the
k-correction (using the {\tt KCORRECT} algorithm of Blanton \& Roweis 2007).
Col:(11). The merger timescale estimated by the formula of Kitzbichler \& White (2008).
Col:(12). The asymmetry factor for the residual image.
Col:(13). $r-$band apparent magnitude of the residual image.
Col:(14). The flux ratio of residual and initial image within 3$R_{\rm
  e}$ (Here $R_{\rm e}$ is
the fitted effective radius from our model).
}
\label{tab:bcg}
\end{table*}

\section{Results}

\subsection{Merger rate and stellar assembly of BCGs \label{sec:assembly}}

We have identified 18 ($\sim$3.5$\%$ of the total) major dry
merging pairs (or triples) involving the BCG in 515 C4 clusters at
0.03 $\leq$ z $\leq$ 0.12 . It has usually been assumed that
physically bound pairs will merge due to dynamical friction. The
merger timescale is, however, uncertain since it depends on a
number of factors, including the separation, relative velocity and
mass ratios of galaxies. Kitzbichler \& White (2008) noticed 
that previous studies on merger rates based on pair statistics have yielded 
a wide variety of results (see their paper for a review). This diversity 
can be attributed to differences in the pair definition and in the timescales adopted.  
They thus used the Millennium Simulation to calibrate the merging timescale as a
function of the stellar mass and projected separation. They find
an average merging timescale (appropriate for our stellar mass and
the Hubble constant): \beqn t_{\rm{merge}} = 2.2\, {\rm Gyr}\,
\frac{\rp}{50\kpc} \left(\frac{M_*}{5.6\times 10^{10}
    M_\odot}\right)^{-0.3} \left(1+\frac{z}{8}\right),
\eeqn
where $M_*$ is the total stellar mass, and $z$ is the (cluster) redshift.
We take a mass-to-light ratio of $M_*/L_r\approx 5 M_\odot/L_\odot$ (Rines et al. 2007;
Patton et al. 2000), yielding a mean total stellar mass of $M_* \sim 8.5
\times 10^{11}M_\odot$ for our massive mergers. 
We estimate the merger timescale for each merger by this formula. 
We find the merger timescale of our mergers ranges from 0.04 to $0.55\Gyr$ 
with a mean value of $0.3\Gyr$ and a standard deviation of $0.17\Gyr$ (see Column 11 in Table~\ref{tab:bcg}).
For comparison, if we apply the formula of Masjedi et al. (2006) to calculate the dynamical 
timescale of our mergers, we obtain a similar timescale of $\sim0.3\Gyr$. 
However, a somewhat shorter dynamical friction timescale of $\sim0.1\Gyr$ is 
obtained if we apply the formula of Patton et al. (2000), although the dynamical 
friction formula may be less applicable to major mergers (Binney \& Tremaine 1987). 
We also note that Bell et al. (2006) obtained a shorter timescale of
$\sim0.15\Gyr$ from simulations for massive early-type mergers.
It seems that the average merging timescale is likely uncertain by a factor of $\sim 2$.

To evaluate the number of mergers BCGs has experienced since (for
example) $z=0.7$, we need to understand how the dry-merger
fraction evolves as a function of redshift. Parameterizing the
evolution of pair fraction in the form of $(1+z)^m$, several
previous studies  (e.g., Le F\`evre et al. 2000; Patton et al.
2002; Conselice et al. 2003; Lin et al. 2004; De Propris et al.
2005; Kartaltepe et al. 2007; Lotz et al. 2008) derive a positive
slope in the range of $m\sim0-3$ for all mergers (involving dry or
wet mergers). Lin et al. (2008), for the first time, estimated the
merger fraction of red galaxies (the majority of them should be
dry-mergers) decreases with redshift slightly with
$m=-0.92\pm0.59$. This suggests that dry-mergers may
preferentially occur in the local Universe. In contrast, a recent
work by Khochfar \& Silk (2008) suggests a nearly constant
dry-merger rate at $z\leq1$.

If we adopt an average merger timescale, and assume a constant
fraction (18/515 $\sim$ 3.5$\%$) of BCGs are in major dry-mergers
over the 6.3 Gyr interval since redshift 0.7, then a present-day
early-type BCG will have experienced $6.3/\tmerge \times 18/515$
$\sim 0.7\,(\tmerge/0.3\Gyr)^{-1}$ major dry-mergers. On the
other hand, if the fraction evolves as $(1+z)^{-0.92\pm0.59}$ (Lin
et al. 2008), then the number of mergers changes to $\sim 0.5
(\tmerge/0.3\Gyr)^{-1}$. Below we adopt an average number of
mergers of these two evolution scenarios,
$0.6\,(\tmerge/0.3\Gyr)^{-1}$. In any case, the uncertainty in the
merger timescale is much larger than that due to the evolution of
the merger rate.

Our merger pairs/triples have an average magnitude difference of
$\sim 0.7$, which corresponds to a $\sim 1:2$ luminosity ratio. In
the merging process, a substantial fraction of the companion
galaxy may be tidally stripped, and so not all the mass of the
companion galaxy will be added to the main one (Yang et al. 2009). We assume a
fraction of $\fmass$ of the companion galaxy is accreted to the
primary galaxy. Thus in each merger, the central galaxy increases
its luminosity (or mass) by $25\%(\fmass/0.5)$. It follows that a
present-day BCG has on average increased its luminosity by $15\%\,
(\tmerge/0.3\Gyr)^{-1}(\fmass/0.5)$ from $z = 0.7$ at a rate of
$2.5\%\,(\tmerge/0.3\Gyr)^{-1}(\fmass/0.5)$ per Gyr. Thus a
non-negligible fraction of the stellar mass of a present-day
early-type BCG is assembled via major dry-mergers since $z=0.7$.

\subsection{Dependence on environments} \label{sec:rich}

\begin{figure}
\centering
\includegraphics[width = 8.1cm]{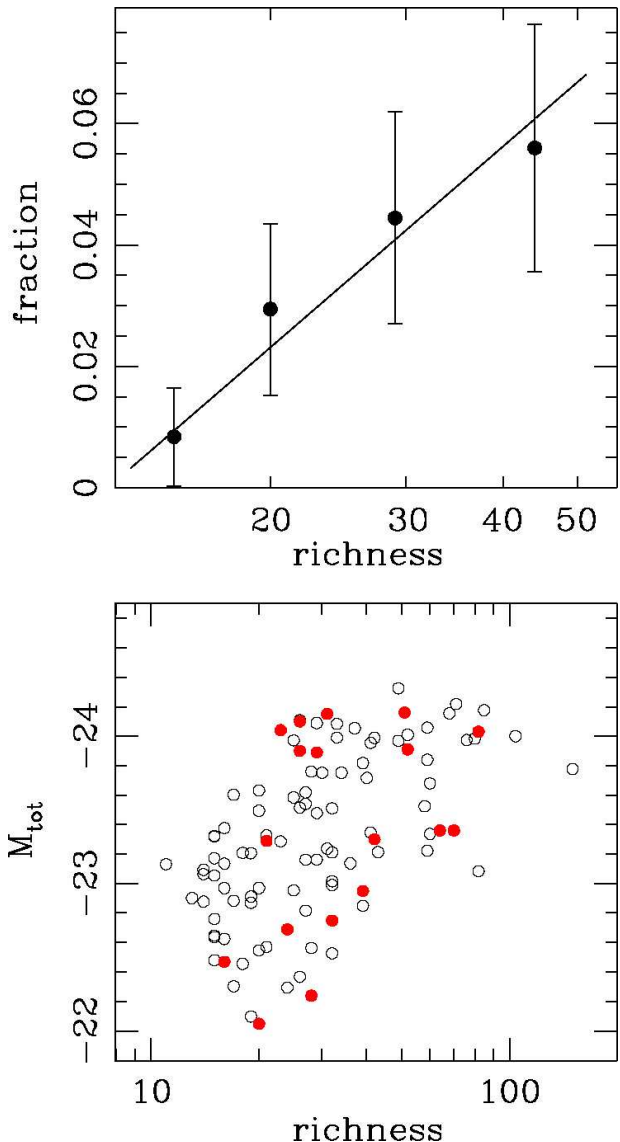}
\caption{The top panel shows
the major dry-merger fraction as a function of the richness of the parent clusters.
Poisson errors are shown. The straight line shows the best 
regression fit ${\rm fraction}=(0.11\pm 0.04) \log {\rm richness} - (0.12\pm 0.05)$. 
The bottom panel shows the BCG luminosities versus the richness for 85 BCGs (open circles)
in Paper I and 18 merging BCGs presented in this work (solid circles).
}
\label{fig:envir}
\end{figure}

In the hierarchical structure formation model, clusters of
galaxies form through mergers of sub-clusters and small groups; in
the process central galaxies are expected to grow together with
clusters. This scenario suggests that the evolution of BCGs may depend
on the environments. We thus examine the relation between the
fraction of BCGs involved in major dry-mergers and the richness of
the parent clusters.

Figure~\ref{fig:envir} shows the results. The fraction of
merging BCGs appears to increase as the clusters become richer. 
A simple linear fit in the form of ${\rm fraction}=A \log {\rm richness} + B$ yields
$A=0.11 \pm 0.04, B=-0.12 \pm 0.05$ (shown as the solid line), and so the significance 
is not very high due to the large error bars. The higher
frequency of more massive dry-mergers in richer clusters will lead
to more luminous (or more massive) BCGs. This is empirically seen 
in observations, as illustrated in the bottom panel of Figure~\ref{fig:envir}.

\section{Summary and Discussion} \label{sec:discussion}

In this work, we have searched for BCGs involved in major
dry-mergers in a sample of early-type BCGs selected carefully from
C4 cluster catalogue (Miller et al. 2005). 18 ($\sim$3.5$\%$)
major dry-mergers involving BCGs have been identified out of 515
clusters at 0.03 $\leq$ z $\leq$ 0.12. These major mergers are selected 
as pairs (or triples) with $r$-band magnitude difference $\dm<1.5$ and 
projected separation $\rp<30$ kpc, and showing signatures of interaction 
in the form of significant asymmetry in residual images.

To double check whether our results are affected by the choice of
the cluster (C4) catalogue (based on DR2 of SDSS), we have also
searched for major dry-mergers involving BCGs in the sample of 625
clusters refined by von der Linden et al. (2007) from the
unpublished DR3 version of C4 clusters. Their sample selects
preferentially early-type BCGs and discards clusters with very few
member galaxies. For this sample, we find 27 (27/625$\sim$4.3\%)
BCGs are involved in major dry-mergers, entirely consistent with
the fraction (3.5\%) found here for the C4 cluster catalogue. 
It should be emphasized that the goal of our method is to identify merging pairs 
with strong interaction features and calculate their true faction in BCGs.
They should be real, physically bound pairs. However, some physical
bound pairs may not yet show strong interaction signatures. Our
estimated fraction should thus be considered as the lower limit of physical
pair fraction in BCGs.

From the identified major-merger candidates, we conclude that a
present-day BCG may have experienced $0.6\,(\tmerge/0.3\Gyr)^{-1}$
major dry-mergers and increased their mass by
$15\%\,(\tmerge/0.3\Gyr)^{-1}(\fmass/0.5)$ on average from  $z =
0.7$. This fraction is comparable to the mass increase predicted
by mostly minor mergers in semi-analytical galaxy formation models
of De Lucia \& Blaizot (2007), if all mass of companion galaxies
are added to BCGs ($\fmass=1$) as De Lucia \& Blaizot (2007)
assumed. However, the mass increase we found here is due to
major-mergers rather than minor majors. Bernardi (2009) concluded that 
the evolution of BCGs in the velocity dispersion-stellar mass relation favors the growth of BCGs 
via many dry minor-mergers. However, the evidence is not very strong
(see her Fig. 8). A more detailed comparison between theory and observations
is desirable.

The major uncertainties in our estimates arise due to the
calibration of the merging timescale and the fraction of the
satellite galaxy that will be added to the central host. For the
former, we used the calibration of Kitzbichler \& White (2008)
derived from the Millennium simulation. However, it remains to be
seen whether their calibration applies well to the dense
environments such as clusters of galaxies. The parameter 
$\fmass$ is also somewhat uncertain. The stellar mass not 
accreted to the central host may form the envelopes in cD galaxies and/or 
the intracluster light (Lin \& Mohr 2004; Rines et al. 2007; Paper I 
and reference therein). We notice, however, that if all the mass 
is assembled into the final BCGs, some of them may be too bright; the 
resulting luminosity function at the bright end may be
inconsistent with the observations which show little evolution at
$L>2.5L_*$ (e.g., Brown et al. 2007; Cool et al. 2008). We
conclude that a substantial fraction of the companion mass must be
stripped, consistent with the presence of envelopes in giant cD
galaxies (e.g., Paper I).

We have also checked the radio properties of BCGs to investigate
whether major mergers trigger the phenomenon of active galactic
nuclei. We cross-correlate our sample with the radio data from
the FIRST survey (Becker, White \& Helfand 1995). There are 94
($\sim$18.3\%) sources with radio emission out of 515 BCGs; these
fractions are consistent with the statistics of Best et al. (2007)
and Croft, de Vries \& Becker (2007). In comparison, there are 3
radio sources ($\sim$16.7\%) out of 18 merging BCGs, i.e., there
is no difference on radio properties between major-merger and
non-major-merger BCGs within the statistical uncertainties.

McIntosh et al. (2008) identified 38 major mergers of red galaxies
from 845 groups at $z\leq$ 0.12, out of which 18 mergers ($\sim
2.1\%$) are between a central galaxy and a satellite galaxy.  Six
(C1000, C1004, C1011, C1060, C1304, C3157) of our 18 mergers are
also in their sample of 38 major mergers; all six are in the top
ten most massive groups among their major mergers. In other words, 
we are probing the most massive end of their SDSS ``groups'' 
in terms of mass range from $10^{13.5}M_\odot$ to $10^{15}M_\odot$.
They find that more massive mergers tend to
appear in richer systems, a conclusion confirmed by our studies. 
In addition, they find that the merger frequency increases as a
function of halo mass, which may explain why their merger fraction
($\sim 2.1\%$) is slightly lower than ours (3.5\%). The mass
increase rate we find for the BCGs is roughly consistent with their
results (2\%-9\% per Gyr under similar assumptions). We conclude that
major dry-mergers in BCGs are not rare events in the local
Universe and they are an important way to assemble luminous BCGs.

\section*{Acknowledgments}

We thank C. N. Hao, J. L. Han for useful discussions, and
in particular A. von der Linden for sharing data and helpful
discussions. We acknowledge the anonymous referee for a 
constructive report that improved the paper. 
This project is supported by the Doctoral Foundation
of SYNU of China (054-55440105020), and by the NSF of China
10833006, 10778622 and 973 Program No. 2007CB815405.
SM acknowledges the Chinese National Science Foundation and
Tianjin Municipal Government for travel support.
Funding for the creation and distribution of the SDSS Archive has
been provided by the Alfred P. Sloan Foundation, the Participating
Institutions, the National Aeronautics and Space Administration,
the National Science Foundation, the U.S. Department of Energy,
the Japanese Monbukagakusho, and the Max Planck Society. The SDSS
Web site is http://www.sdss.org/. The SDSS is managed by the
Astrophysical Research Consortium (ARC) for the Participating
Institutions. The Participating Institutions are The University of
Chicago, Fermilab, the Institute for Advanced Study, the Japan
Participation Group, The Johns Hopkins University, the Korean
Scientist Group, Los Alamos National Laboratory, the
Max-Planck-Institute for Astronomy (MPIA), the
Max-Planck-Institute for Astrophysics (MPA), New Mexico State
University, University of Pittsburgh, Princeton University, the
United States Naval Observatory, and the University of Washington.

\label{lastpage}
\end{document}